\documentclass[11pt,twoside
]{article}

\usepackage{baaa2010}
\usepackage{graphicx}
\usepackage{subfigure}
\usepackage{psfrag}
\usepackage{amssymb}
\usepackage[spanish,activeacute,english]{babel}
\usepackage[latin1]{inputenc}
\usepackage[T1]{fontenc} % Computer Modern (CM) fonts
\usepackage{ae,aecompl} % and: dvips -Pcmz -Pamz macros_aaa.dvi
\usepackage{latexsym}
\usepackage{verbatim}
\usepackage{amsmath}
\usepackage{stmaryrd}
\usepackage{amsfonts}
\usepackage{amssymb}
\usepackage{wasysym}
\usepackage[colorlinks=true,dvips]{hyperref}
\begin{document}
%%%%%%%%%%%%%%%%%%%%%%%%%%%%%%%%%%%%%%%%%%%%%%%%%%%%%%%%%%%%%%%%%%%%%%%%%%
%%%% SELECCIONE EL IDIOMA EN QUE SE ESCRIBE EL ART?CULO:              %%%%
%\myselectspanish

\myselectenglish
%%%%%%%%%%%%%%%%%%%%%%%%%%%%%%%%%%%%%%%%%%%%%%%%%%%%%%%%%%%%%%%%%%%%%%%%%%
\vskip 1.0cm 
\markboth{Noelia Jim\'enez et al.}
 {Chemical effects on the CMR}
\pagestyle{myheadings}
%%%% DESCOMENTE LA LINEA QUE DESCRIBE EL CARACTER DE SU TRABAJO       %%%%
\vspace*{0.5cm}
%\noindent TRABAJO INVITADO 
\noindent PRESENTACI\'ON ORAL
%\noindent PRESENTACI\'ON MURAL
%\noindent RESUMEN 
\vskip 0.3cm
\title{Chemical effects on the development of the colour--magnitude relation of cluster galaxies}

%\title{ Template paper for publication in the Bulletin of the 
%Argentinian Astronomical Association with instructions for the use of 
%\LaTeX{}}

\author{ Noelia Jim\'enez$^{1,2,3}$, Sof\'ia A. Cora$^{1,2,3}$, Lilia P. Bassino$^{1,2,3}$, Anal\'ia Smith Castelli$^{1,2,3}$, Tom\'as Tecce$^{2,4}$}

\affil{ (1) Facultad de Ciencias Astron\'omicas y
  Geof\'isicas(FCAG-UNLP)\\ (2) Consejo Nacional de Investigaciones
  Cient\'ificas y T\'ecnicas(CONICET)\\ (3) Instituto de Astrof\'isica
  de La Plata (CCT La Plata,CONICET)\\(4) Instituto de Astronom\'a y
  F\'sica del Espacio (IAFE), Buenos Aires, Argentina}
\begin{abstract}
We investigate the development of the colour-magnitude relation (CMR)
of cluster galaxies. This study is carried out using a semi-analytic
model of galaxy formation and evolution coupled to a sample of
simulated galaxy clusters of different masses, reinforcing the
conclusions reached by Jim\'enez et al. (2009).  We compare both
simulated and observed CMRs in different colour-magnitude planes,
finding a very good agreement in all cases. This indicates that model
parameters are correctly tuned, giving accurate values of the main
properties of galaxies for further use in our study. In the present
work, we perform a statistical analysis of the relative contribution
to the stellar mass and metallicity of galaxies along the CMR by the
different processes involved in their formation and evolution
(i.e. quiescent star formation, disc instability events and galaxy
mergers). Our results show that a mix of minor and major dry mergers
at low redshifts is relevant in the evolution of the most luminous
galaxies in the CMR. These processes contribute with
low metallicity stars to the remnant galaxies, 
thus increasing the galaxy masses without
significantly altering their colours. These results are found for all
simulated clusters, supporting the idea of the universality of the CMR
in agreement with observational results.
\end{abstract}

\begin{resumen}
Investigamos el desarrollo de la relaci\'on color--magnitud (RCM) de
las galaxias residentes en c\'umulos. El estudio se realiza combinando
un modelo
semianal\'itico de formaci\'on de galaxias con
simulaciones cosmol\'ogicas de {\em N}-cuerpos de c\'umulos de galaxias
de distintas masas. 
Los resultados obtenidos refuerzan las conclusiones dadas 
en Jim\'enez et al. (2009). 
Comparamos las RCM simuladas con las 
observadas en tres planos color--magnitud, obteniendo muy buen acuerdo en 
los tres casos, lo que indica 
que los par\'ametros del modelo est\'an bien calibrados y los valores 
obtenidos de las principales propiedades de las galaxias son adecuados 
para el uso subsiguiente. En este trabajo, realizamos un an\'alisis 
estad\'istico de la contribuci\'on relativa a la masa estelar y a la 
metalicidad 
de las galaxias en la RCM, de los procesos de formaci\'on y evoluci\'on sufridos
por estas galaxias (formaci\'on estelar pasiva, inestabilidad de disco
y fusiones de galaxias). Nuestros resultados
muestran que una mezcla de fusiones menores y mayores de tipo
seco (con poco contenido de gas), es determinante en la evoluci\'on 
de las galaxias
m\'as luminosas de la RCM, ya que agregan material estelar de baja
metalicidad a las galaxias remanentes, de forma tal que su masa crece
pero los colores no se ven afectados significativamente. Los mismos
resultados han sido hallados en todos los c\'umulos simulados, apoyando
la idea de la universalidad de la RCM, en concordancia con
trabajos observacionales.

\end{resumen}
\section{Introduction}
The colour--magnitude relation (CMR) is usually understood as a
mass--metallicity relation: more luminous and massive galaxies in this
relation have deep potential wells capable of retaining the metal
content released by supernovae events and stellar winds. Generally, a
linear relation has been used to fit the correlation between
luminosity and colour of cluster galaxies lying in the CMR.  However,
different fits have been suggested (e.g. Janz \& Lisker 2009),
consistent with a change of slope from the bright to the faint
ends. Additional evidence of a tilt towards bluer colours at the
bright end of the CMR arises from studies of large samples of
early-type galaxies in the Sloan Digital Sky Survey (Baldry et
al. 2006, Skelton et al. 2009). This observed trend has been
reproduced by Skelton et al. (2009) using a simplified model in which
dry mergers of galaxies already on the CMR mildly change the slope of
this relation at higher luminosities.  This detachment of the bright
end of the CMR, motivates our study.  We use a semi-analytic model of
galaxy formation and evolution that includes the effect of feedback
active galactic nuclei (Lagos et al. 2008), and an improved estimation
of the scalelength of galactic discs (Tecce et al. 2010).  Our aim is
to explain how minor mergers influence the evolution of the most
massive galaxies in the CMR, since several authors consider dry
mergers as the prime candidates to account for the strong mass and
size evolution of the stellar spheroids at $ z<2$, and they are
supposed to increase the stellar mass of galaxies, without changing
the colours (Bernardi et al. 2007).

\section{Comparison between simulated and observed} 

 We extend the study performed by Jim\'enez et al. (2009), considering
 two sets of simulated galaxy clusters, C14 and C15, with virial
 masses in the ranges $\simeq (1.1-1.2)\times 10^{14}\,h^{-1}\,{\rm
   M}_\odot$ and $\simeq (1.3-2.3)\times 10^{15}\,h^{-1}\,{\rm
   M}_\odot$, respectively (see Dolag et al. 2005 for details), and
 construct the CMRs in three different colour--magnitude
 planes. Minimum square fits to the simulated CMRs are estimated in
 each magnitude system, for all clusters. The average slopes
of these fits are $b_{\rm (B-R)}= -0.0426$ in
 the $(B-R)$ vs. $R$ plane, $b_{\rm (V-I)}= -0.035$ in the
 $(V-I)$ vs. $I$ plane ($R$ and $I$ magnitudes are in the Cousins
system), and $b_{\rm (C-T_{\rm 1})}= -0.0740$ 
 in the $(C-T_{\rm 1})$ 
 vs. $T_{\rm 1}$ plane (Washington photometric system). We compare 
these slopes with observed ones.  In the $(B-R)$ vs. $R$ plane, 
L\'opez-Cruz et al. (2004) studied the
 CMRs of 57 low-redshift cluster galaxies, obtaining an average slope of
 $b_{\rm (B-R)}= -0.051 \pm 0.002$. 
For the $(C-T_{\rm 1})$ vs. $T_{\rm 1}$ plane, Smith Castelli et
al. (2008), obtained the CMR of early-type galaxies in the Antlia
cluster using the Washington photometric system. This CMR is
characterized by  $b_{\rm (C-T_{\rm 1})}=-0.073 \pm 0.005$.  These
slopes, as well as those observed in the $(V-I)$ vs.  $V$ plane, are
in concordance with our findings (see Jim\'enez et al., in prep.)
allowing us to conclude that the model reproduces very well the CMRs
in different colour--magnitude planes.

\section{Physical processes involved in the development of the CMR}

We find an interesting aspect from our simulations. In all
colour--magnitude planes, the most massive and luminous galaxies
present a detachment from the general
trend denoted by fainter galaxies.  This break occurs at approximately
the same magnitude in the different filters ($M_R^{\rm break} \sim
M_V^{\rm break} \sim M_{T1}^{\rm break} \approx -20$), being more
evident in the Washington T1. In order to 
understand the physical reasons that lead to the bluer colours of the brightest
galaxies in our simulations, we divide the simulated CMR at $z=0$ in bins of
one magnitude from $M_{\rm T_{\rm 1}} =-17$ to $M_{\rm T_{\rm 1}}
=-24$ and analyze the evolution of mass and metallicity of the stellar
component contributed by different processes: quiescent star formation (QSF)
and starbursts during major wet, major dry, minor wet and minor dry
mergers, and disc instability (DI) events (for details on the classification
of these processes in our model, see Jim\'enez et al. 2009). 

Figure 1 shows the evolution with redshift of the accumulated mass
contributed by different sets of processes (QSF and
DI: left panel; minor mergers: middle panel)
normalized to the
total stellar mass acquired within each magnitude bin. In this way,
we have an estimation of the relative importance of these processes 
on the formation of galaxies with different luminosities. 
\begin{figure*}

\includegraphics[width=0.32\textwidth]{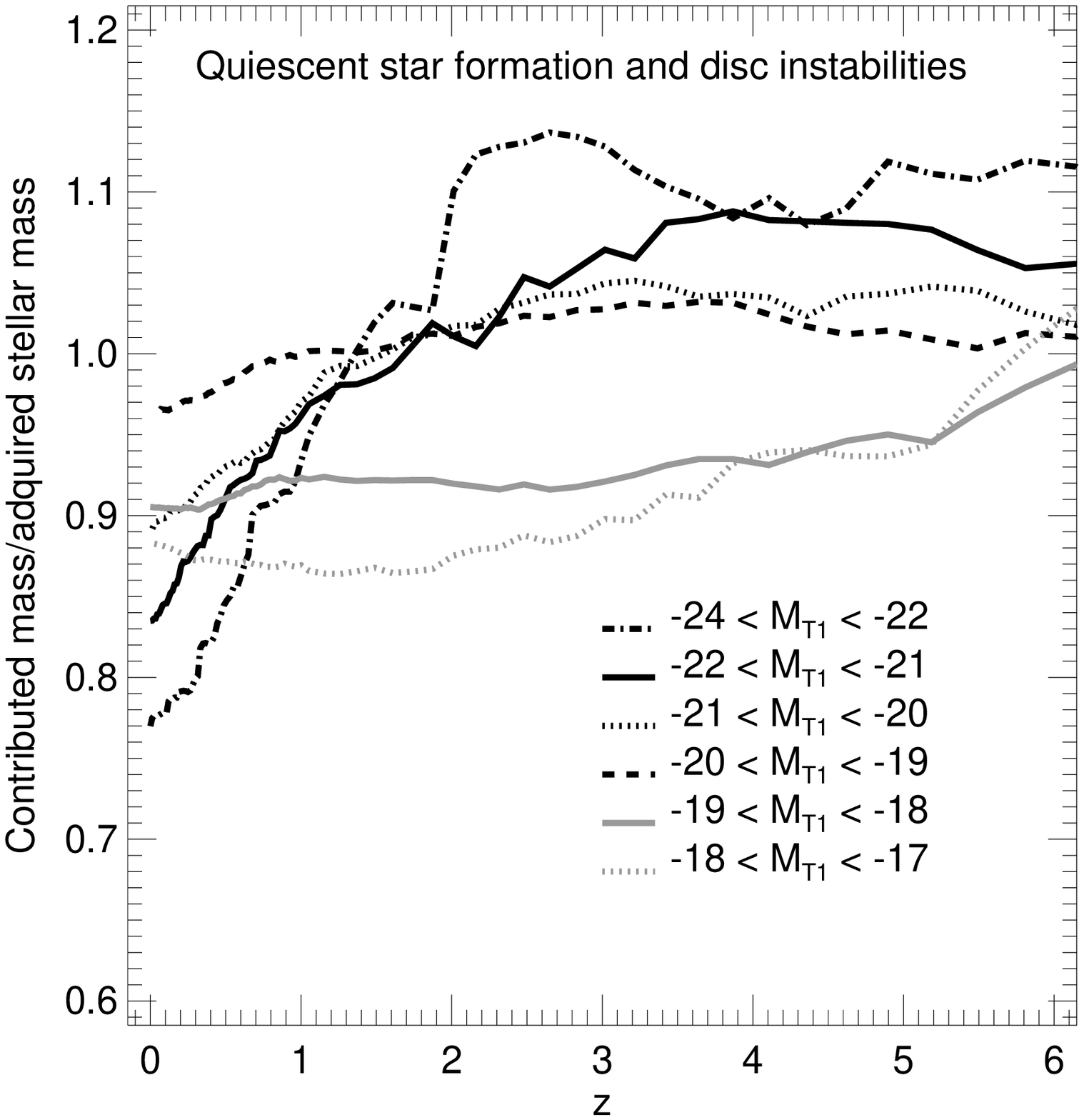}
\includegraphics[width=0.32\textwidth]{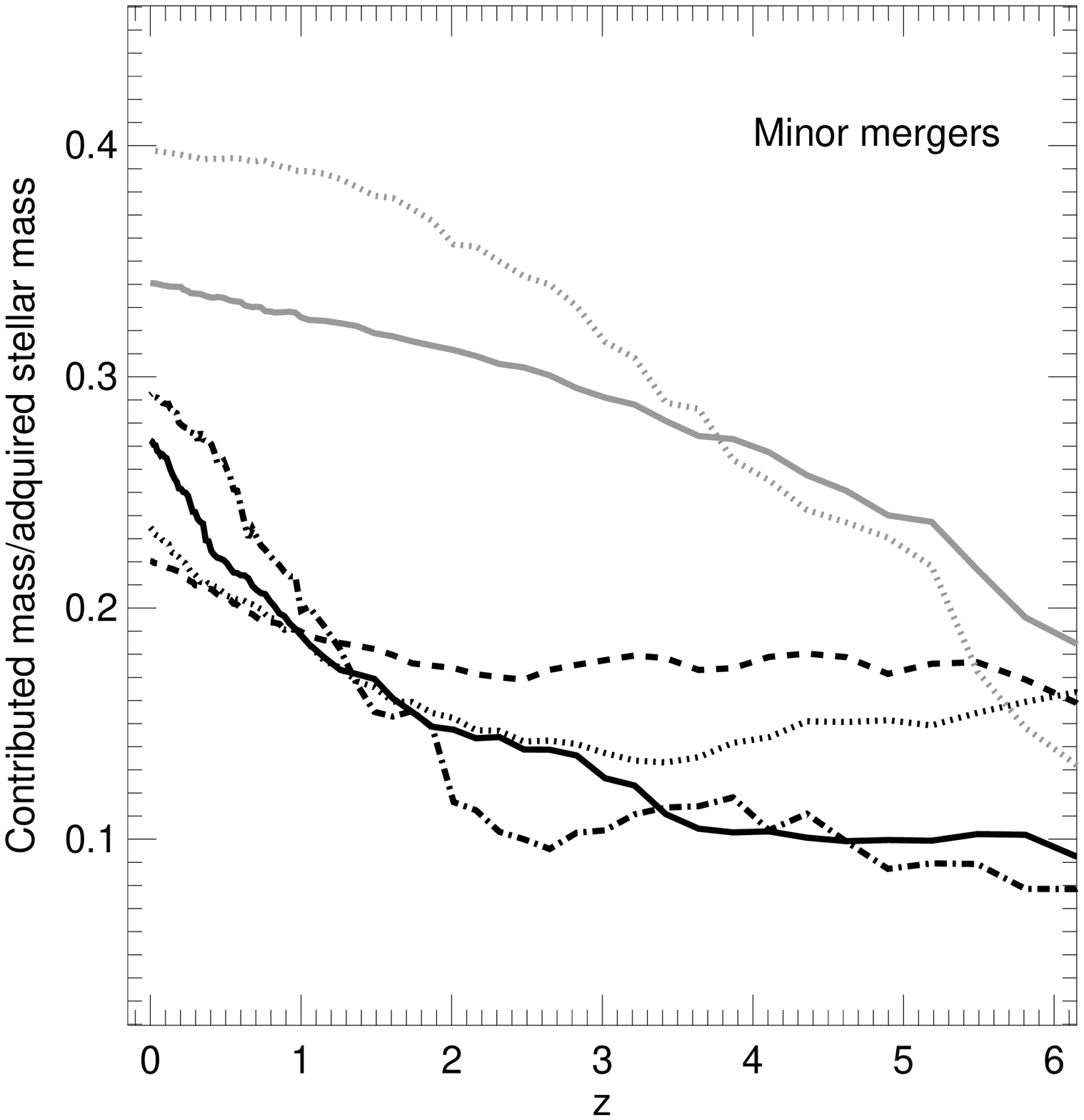}
\includegraphics[width=0.32\textwidth]{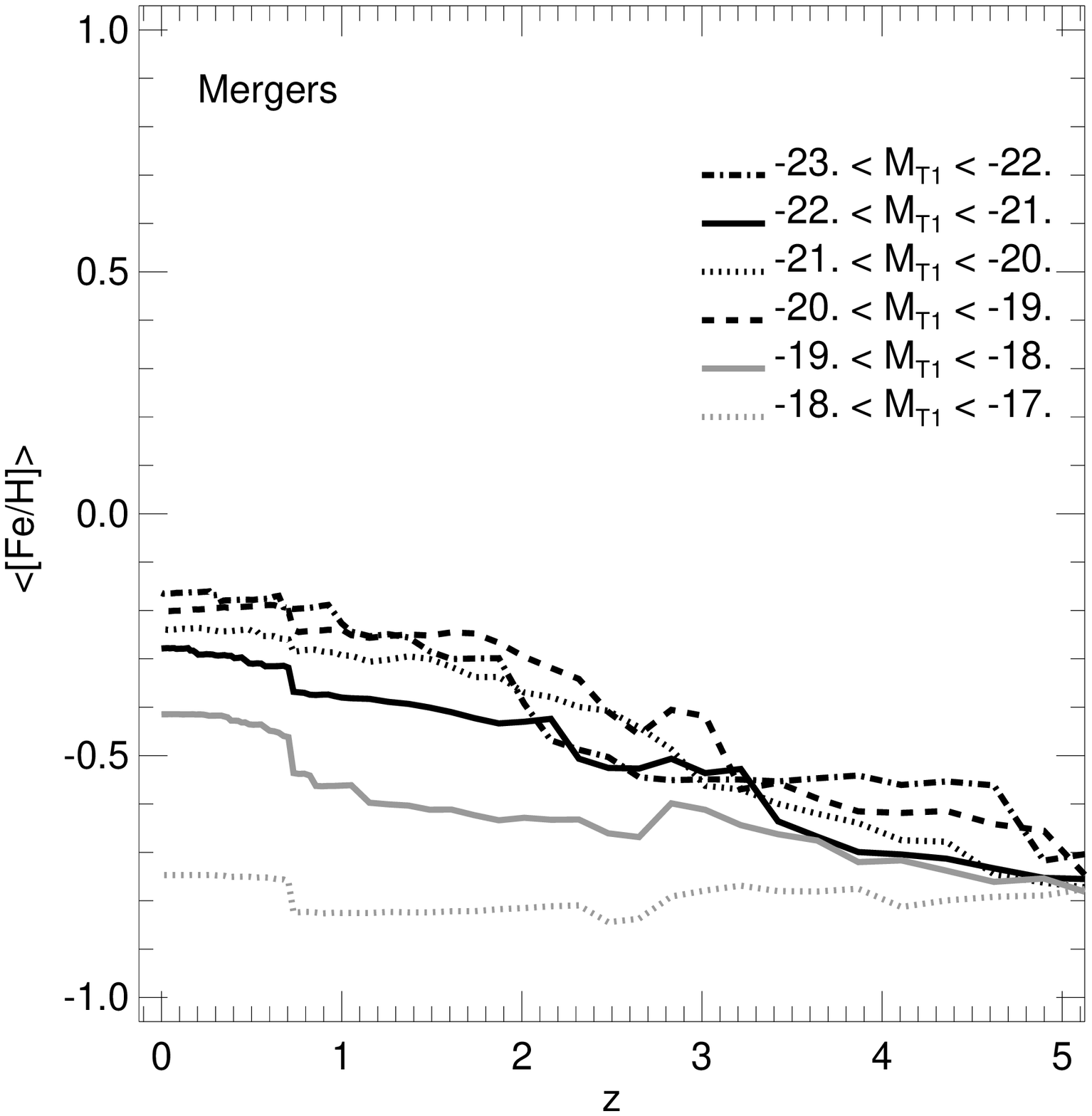}
\caption{{\em Left panel}: Evolution of stellar mass contributions by QSF 
  and DI relative to the total mass acquired by
  galaxies within different magnitude bins. {\em Middle panel}: Idem to left 
panel but for merger events. {\em Right panel}: Evolution of the
  metallicity of the stellar mass contributed by merger
  events.}
\label{merTotDM}
\end{figure*}
It can be seen that for a
set of very luminous galaxies ($ -24 \leq M_{\rm T_{\rm 1}} \leq
-20$), the mass fraction contributed by QSF and DI decreases at low redshifts
(from $z \approx 2 - 3$) as the galaxy is being formed.
Note that values larger than unity are reached because the total
stellar mass of galaxies is
reduced as a result of mass recycled by 
the stellar population due to mass loss and dying stars, 
following the whole evolution traced by the semianalytic code;
this is not taken into
account by the individual mass contributions. 
As the stellar mass of galaxies increase, 
QSF and DI lose importance at expense of the effect of mergers.
In particular, the role of minor mergers becomes relevant, as
it is evident by the fractions presented in the middle panel of Fig. 1, 
characterized by a steep increase since  $z \approx 2 - 3$.
This increment is less pronounced for less luminous galaxies (grey lines),
although the relative contribution of this process to the galaxy mass
already acquired is larger than for brighter galaxies.
Besides, these fainter galaxies keep forming stars through QSF and DI
as indicated by the almost constant values depicted by grey lines since
$z\approx 3$ (left panel), being consistent with the fact that
the stellar mass keep growing as it consumes the available reservoir of 
cold gas. 
As shown by Jim\'enez et al. (2009), minor mergers involved in the evolution
of the brightest galaxies are mainly dry mergers (fraction of gas cold gas
content in the central galaxy less than 0.6). 
Major dry mergers also affect their formation but are less important in terms
of the contributed mass fraction.
On the other hand, fainter galaxies also suffer SF associated to starbursts 
occurring during minor wet mergers.

The total metallicity achieved at $z=0$ by the most luminous
galaxies in the CMR, considering all physical processes, lies within the
range $-0.2 \leq \rm{[Fe/H]_{\rm Total}} \leq 0.15$ (not
shown here). 
Taking into account that mergers (mainly minor dry ones) play a significant
role in the evolution of these galaxies,
we explore the evolution of the metallicity of the stars accreted and 
formed during 
these events. This is shown in the right panel of Figure $1$,
where different line types depict the results obtained for galaxies
in different magnitude bins. At $z=0$, the metallicity contributed by these
components to the most luminous galaxies ranges within
$ -0.4 \leq \rm{[Fe/H]_{\rm Mergers}}\leq -0.15 $.  
Then, accreted stars in merger
events contribute with metals having subsolar abundances. Hence
minor and major dry mergers 
suffered by the most luminous galaxies of the CMR since $z <2$
lead to the increase of their stellar mass
without strongly affecting their metallicities. Thus a break
in the bright end of the CMR arises with galaxies
rising their total luminosities but without changing their
colours. 
The same results are found for $C14$ and $C15$ clusters,
suggesting that the CMR is universal, in concordance with 
previous observational works (e.g. L\'opez Cruz et al. 2004).

\begin{referencias}
\reference Bernardi, M., Hyde, J. B., Ravi, S. K., Miller, J. C.,
Nichol, R. C., 2007, \apj, 133, 1741 \reference Baldry, I. K.,
Balogh,M. L., Bower, R. G., et al., 2006, MNRAS, 373, 469 \reference
Dolag, K., Vazza, F., Brunetti, G., Tormen, G., 2005, MNRAS, 364, 753
\reference Janz, J., Lisker, T., 2009, \apj, 696, 102L \reference
Jim\'enez, N., Cora, S .A., Bassino L. P., Smith Castelli, A., 2009,
BAAA, 52, 197 \reference Lagos, C., Cora, S. A., \& Padilla N. D.,
2008, \mnras, 388, 587 \reference L\'opez-Cruz O., Barkhouse W. A.,
Yee H. K. C., 2004, ApJ, 614, 679 \reference Skelton, R. E. , Bell,
E. F., Somerville, R. S. 2009, \apj, 699, 9L \reference Smith
Castelli, A. V., Bassino, L. P., Richtler T., et al., 2008, MNRAS,
386, 2311 \reference Tecce T. E., Cora S. A., Tissera P. B., Abadi
M. G., Lagos C. del P., 2010, MNRAS, 408, 2008
\end{referencias}
\end{document}